# APPLYING COMPLEXITY THEORY TO A DYNAMICAL PROCESS MODEL OF THE DEVELOPMENT OF PATHOLOGICAL BELIEF SYSTEMS

**Brian P. O'Connor and Liane Gabora**
Department of Psychology, University of British Columbia,
Kelowna, BC, CANADA

**ABSTRACT**
A general dynamical process model of psychiatric disorders is proposed that specifies the basic cognitive processes involved in the transition from beliefs about self, others and world that are normal and adaptive, to beliefs that are rigid, extreme, and maladaptive. The relevant thought trajectories are self-confirming, and are considered to underlie the corresponding trajectories in symptoms. In contrast with previous work, the model focuses on underlying mechanisms, and it provides an evolutionary basis for the widespread susceptibility to psychiatric symptoms and disorders without the problematic claim that such disorders were selected by evolutionary forces. The model thereby incorporates both normality and abnormality in the same framework.

## INTRODUCTION

Humans are susceptible to a wide variety of mental disorders that, despite their substantial differences, are all characterized by distinctive, rigid, extreme beliefs about themselves and others, and maladaptive ways of acting in the world. Examples of such beliefs include "I am a failure," "I am incapable of having positive interactions with others," and "I am overweight and cannot believe those who tell me I am dangerously thin." The maladaptive beliefs associated with psychiatric disorders are evident and challenging to anyone who interacts with the individuals on a regular basis, including family members and mental health professionals. Yet most people were not born with such peculiar beliefs, and their family members can recall times when the client did not have such problematic views or any other psychiatric symptoms. How did they come about?

We propose that they are side-effects of the fact that the web of understandings a human mind weaves about the world is not just complex, dynamic, self-organizing (Orsucci, 2002) but also self-regenerating, self-perpetuating, and thereby 'alive', and evolving not just at the biological organismic level but at a second, cultural level (Gabora, 1998, 2004, 2008). The ability to weave an internal model of the world, or worldview, stems from the specifically human capacity for a self-triggered stream of thought, in which one thought triggers another, which triggers another, and so forth; our daily experience is formed not just by sensory impressions but also by mental operations on these impressions. These mental operations have the desirable effect of structuring impressions into a more-or-less coherent web of understanding, the integrated nature of which enables us to plan and prioritize, draw analogies, adapt behavior to the specifics of situations, and so forth. However, the proclivity to mentally operate on impressions also provides for the possibility that they become increasingly distorted, particularly if the developing worldview that shapes this assimilation process is becoming a biased model of the external world.

The goal of this chapter is to unpack this argument and propose a tentative, general model of the process by which once-normal individuals develop deep, maladaptive convictions that are seemingly impervious to refuting evidence and that are closely intertwined with other symptoms. Although the contents of beliefs vary across disorders, it is proposed that the basic cognitive processes involved are essentially the same across disorders. The model incorporates the possible influences of affective and biological factors on the trajectories of belief systems. The focus is on cognition, but there is no claim that cognitive factors are always primary or causal. Distinct belief systems are nevertheless almost always present in psychiatric disorders, and we are simply proposing a general model of how they develop. A proper understanding of how such belief systems develop will presumably be important to understanding how rigid belief systems can change for the better.

The chapter begins with descriptions of the particular beliefs that are commonly involved in a variety of Axis I and Axis II disorders from the Diagnostic and Statistical Manual (DSM-IV-TR; American Psychiatric Association, APA, 2000). In the second section, we provide an outline of relevant concepts from dynamical systems and complexity theory. Distinctions are made between how the present model differs from previous applications of these concepts to psychopathology. We also describe research on memory and thought processes, and on interpersonal processes, that are involved in the trajectories of belief systems. The fourth section describes the implications of our model and the corresponding answers to a number of perennial questions in the literature.

## PROTOTYPICAL BELIEFS IN A VARIETY OF PSYCHIATRIC DISORDERS

We begin by outlining the primary beliefs associated with particular disorders (Alloy and Riskind, 2006; Reinecke and Clark, 2004; Riso, du Toit, Stein, and Young, 2007). Our focus is on the beliefs *per se* and we do not provide full descriptions of the various disorders, which can be found in numerous other sources, including the DSM-IV-TR (APA, 2000). We also do not review the research on all of the cognitive phenomena associated with each disorder. Instead, we simply summarize prototypical belief constellations. We hope readers will notice the differences and similarities in the beliefs across disorders, and the fact that a number of beliefs are associated with each disorder. The beliefs also vary greatly in how general versus specific they are. Some pertain to very general thoughts about oneself (e.g., "I am a bad person"), whereas others are much more focused and specific (e.g., "My heart is racing and I am about to die").

**Depression**
Depression typically involves an assortment of unrealistic, negative beliefs that can sometimes be delusional (Hollon and DeRubeis, 2004; Derubeis, Young, and Dahlsgaard, 1998). There are beliefs about the self (e.g., "I'm worthless"), about past events (e.g., "The breakup with my girlfriend was all my fault"), about the future ("I'll never get another job"), and about the world ("People are uncaring"). Automatic negative thoughts in specific situations ("I can't do this") are generated by more general and abstract underlying beliefs and assumptions ("I am incompetent and unlovable"). Depressive schemas or worldviews are believed to precede and cause depressive episodes. According to the hopelessness model of depression (Abramson, Metalsky, and Alloy, 1989), vulnerability to depression is increased by a depressogenic inferential style. This involves beliefs that negative or stressful events are caused by internal, stable, and global factors. The resulting negative expectations or hopelessness then generates the other symptoms of depression. Numerous measures of the beliefs that commonly occur in depression have been developed (Hollon and DeRubeis, 2004), including a measure of the delusional beliefs that occur in psychotic major depression (Meyers et al., 2006). There is even a measure of the beliefs that depressed persons tend to have about their depression (Thwaites, Dagnon, Huey, and Addis, 2004), which is important because such beliefs affect the perceived credibility of treatments and therapy outcomes.

**Mania**
Inflated self-beliefs (self-esteem, grandiosity) are defining features of mania. Beliefs about one's knowledge, achievements, interpersonal relationships, and abilities are exaggerated, as are perceptions of what one can accomplish. Newman and Beck (1992) describe a positive cognitive triad model of mania in which the self is seen as highly valued and powerful, experiences are viewed as overly positive, and the future is believed to be full of opportunities. Experiences of grandiosity are perpetuated by magnification of positive feedback, and minimization of, or obliviousness to, negative feedback. On the other hand, mania can also involve paranoid thinking, which is reinforced by selective attention to evidence that confirms paranoid beliefs, along with dismissal of disconfirming evidence (Derubeis, Young, and Dahlsgaard, 1998). Measures have been developed that assess the beliefs that commonly occur in mania and bipolar disorder (Beck, Colis, Steer, Madrak, and Goldberg, 2006; Mansell, Rigby, Tai, and Lowe, 2008). There is also a recent theory that describes how mood swings may be partially caused by the beliefs and interpretations that occur in manic and depressive states (Mansell, Morrison, Reid, Lowens, and Tai, 2007).

**Panic Disorder**
The primary beliefs in panic attacks involve catastrophic misinterpretations of bodily sensations that cause spiraling anxiety and autonomic responses that, in turn, provide further evidence for the alarming interpretations (Salkovskis, 1998). Particular bodily sensations are interpreted as signs of imminent personal disasters. "I'm having a heart attack", "I'm dying", "I am losing my mind." "I'm losing control over my own behavior." Pre-existing experiences, assumptions, and beliefs about anxiety make some patients prone to such expectations for even minor triggering events or situations, which culminate in full-blown attacks within seconds. Physiological symptoms generated by minor daily stresses or even caffeine can result in being terrified and convinced of the existence of profound threats to their personal safety. Wenzel, Sharp, Brown, Greenberg, and Beck (2006) developed a four-factor measure of the beliefs that commonly occur in panic disorder.

**Social Phobia**
Social phobia involves both general and specific beliefs about oneself and others (Juster and Heimberg, 1998; Ledley, Erwin, and Heimberg, 2008). There are often general beliefs in the necessity of being loved and approved by everyone and in being highly competent in many life domains in order to be worthwhile. Other people are expected to be critical, and social situations are perceived as dangerous and threatening to one's self-esteem. Social phobics tend to believe that perfect social performances are required to avoid social criticisms and that they are unlikely to live up to such high standards. These beliefs result in social phobics, "spectatoring," and routinely monitoring their behaviors in the presence of others and others' evaluations of them. They compare their behavior with the high standards that they believe others hold for them. The social environment (e.g., the nonverbal behaviors of others) is monitored for signs of threat, and internal experiences of physiological arousal are interpreted as signs of failure and looming danger. Public self-consciousness and internal distractions increase, and there may be fears of blushing or appearing ridiculous. This sequence of events strengthens the negative beliefs that social phobics have about their ability to make positive impressions on others. These interpretive tendencies gradually become more powerful and important than the actual reactions and feedback from others (e.g., Abbott and Rapee, 2004). Social inhibition and withdrawal typically follow. Measures of the beliefs that commonly occur in social phobia include those developed by Telch, Lucas, Smits, Powers, Heimberg, and Hart (2004) and by Turner, Johnson, Beidel, Heiser, and Lydiard (2003).

**Generalized Anxiety Disorder (GAD)**
GAD is characterized by excessive and uncontrollable worry about a variety of possible threats. GAD individuals constantly search for signs of possible threats in order to cope with or avoid catastrophes (Beck and Emery, 1985). While some degree of worry is normal, especially given that human thought may have evolved as an adaptive mechanism for anticipating and avoiding dangers, the pervasiveness of worry in GAD clients has been perplexing to clinicians. The beliefs about worrying itself can be both negative (e.g., "My worrying is uncontrollable and harmful to my health") and positive beliefs (e.g., "Worrying helps me cope").

Worry helps individuals avoid processing more painful emotional material, and the cognitive processes involved tend to be self-fulfilling. GAD clients display a pre-attentive, or even unconscious bias to threat cues. They display rapid cognitive avoidance of detected threats, which impairs explicit memory while simultaneously improving implicit memory for the material. They also make negative interpretations and predictions from ambiguous and neutral information. Aversive images are avoided in the short term, but anxious interpretations and beliefs are maintained because corrective information is not properly processed. Worrying thus increases cognitive rigidity (Borkovec and Newman, 1998). Wells (2006) developed a three-dimensional belief-based measure of worry that differentiates GAD patients from patients with other anxiety disorders.

**Obsessive-Compulsive Disorder (OCD)**
A surprisingly wide variety of maladaptive beliefs have been found to occur in OCD, including overestimation of threat and risk, exaggerated responsibility and guilt, beliefs about the need to control thoughts, beliefs about close links between thoughts and actions, perfectionism, and uncertainty and doubt (Obsessive Compulsive Cognitions Working Group, 1997; Wilhelm and Steketee, 2006). Intrusive thoughts are given excessive importance, and there is an exaggerated sense of responsibility for ominous harm coming to self or others. Individuals with OCD also tend to believe that they have the power necessary to prevent negative outcomes.

Some maladaptive beliefs in OCD appear to stem from other maladaptive beliefs. For example, there tend to be: (a) beliefs about the need for perfect competence in order to feel worthwhile; (b) the belief that mistakes or falling short of one's ideals should result in punishment; (c) the belief that one has the power to prevent negative outcomes by performing special rituals; and (d) beliefs that certain thoughts and feelings are unacceptable, potentially catastrophic, and worthy of punishment. These beliefs all contribute to excessively high expectations of threat and aversive outcomes. Similarly, the belief that every situation has a perfect solution and that such perfect solutions must be found results in chronic indecisiveness, discomfort, and doubts about one's experiences and actions. Ritualistic behaviors are then performed to reduce aversive uncertainties.

Maladaptive beliefs also contribute to the intrusive thoughts commonly experienced by OCD patients. Intrusive thoughts are experienced by over 90% of the general population, and they are typically ignored. But they are given excessive importance, attention, and faulty interpretations by those with OCD (Salkovskis, 1989), who tend to believe that simply having an unwanted or objectionable thought is morally equivalent to engaging in the corresponding objectionable behavior (Rachman, Thordarson, Shafran, and Woody, 1995). This "thought-action fusion" contributes to the belief that the imagined negative events are more likely to actually occur. OCD individuals then make efforts to suppress such thoughts, which results in further sensitization and vigilance to all thoughts, causing the cycle of attention to intrusions and attempts at suppression to escalate. OCD individuals eventually believe they must exert complete control over their intrusive thoughts and impulses (Steketee and Frost, 1998). A three-factor measure of the beliefs that commonly occur in OCD was described by Steketee (2005).

**Post-Traumatic Stress Disorder (PTSD)**
The beliefs of people with PTSD involve themes of danger, guilt, shame, alienation, mistrust, betrayal, disillusionment with authority, loss of personal control, and beliefs in one's inability to cope with stressful events (Dalgleish, 2004). The traumatic experiences are realities that cannot be assimilated with existing, normal, preferred beliefs, such as assumptions that people are benevolent, that the world is meaningful, and that oneself is worthy and valuable. The contents of beliefs systems in PTSD vary depending on the kinds of traumas experienced, and a number of measures have been developed to identify the dysfunctional thoughts and distortions that can occur in PTSD cases (Taylor, 2004).

**Eating Disorders**
Individuals with anorexia nervosa continually believe that they are too fat and their desire to lose weight dominates their thoughts and lives. Neither the idea of being too fat nor the fear of gaining weight disappear as the person loses weight. Instead, the belief that one is fat becomes stronger. The person also typically denies the seriousness of the loss of weight. These beliefs are so rigid that experts believe that anorexic patient must have distorted perceptions of their bodies (Fairburn, 2008).

Patients with bulimia nervosa also believe they are heavy, slow, and fat, and they have dysfunctional thoughts about food and eating. Foods are either "good" or "bad". This strict, dichotomous thinking style increases the probability that rules for foods and eating will be broken. Consumption of even a small amount of forbidden food results in the belief that one's diet is "totally broken". Further dysfunctional automatic thoughts follow (e.g., "since my day is already ruined, I might as well continue to eat"). Thus, a minor dietary transgression may culminate in a binge episode.

Those with eating disorders place excessive importance on outward appearance and weight in their general self-evaluations. They believe that to be fat is to be worthless, unlikable, and disgusting, and that to be attractive, successful, intelligent, and to be happy it is necessary to be skinny. These beliefs are baffling because the patients are typically not objectively overweight (Fairburn, 2008; Jansen, 1998). Eating disorder patients commonly trace their negative self-beliefs to trauma or abuse in childhood and they tend to believe that dieting is an effective method of counteracting the negative implications associated with their self-beliefs (Cooper, Todd, and Wells, 1998). Measures of the beliefs that are common among patients with eating disorders were provided by Hinrichsen , Garry, and Waller (2006) and by Young, Klosko, and Weishaar (2003).

**Schizophrenia**
There is an extensive literature on a variety of cognitive impairments that occur in schizophrenia. The present focus will not be on the long list of such impairments but on the beliefs that schizophrenics typically have about themselves and others. Perhaps most obvious are delusions. Beck and Rector (2005) describe how delusional belief systems emerge from biased information processing. Particular schemas become hypersalient, rigid, relatively impermeable to ordinary corrective feedback, and shape perceptions of the world to the exclusion of consensually shared meanings. The end results are firm but clearly false beliefs that someone or something controls their thoughts or plots against them, or that they have special relationships with God. Beck (2004) described how negative beliefs about interpersonal attachments and about one's abilities in schizophrenia provide a basis for persecutory delusions and for negative symptoms. He also described how delusions of grandiosity develop from feelings of loneliness, inferiority, and vulnerability. Measures of the beliefs that sometimes occur in schizophrenia were described by Beck, Baruch, Balter, Steer, and Warman (2004), Csipke and Kinderman (2006), and by Eisen, Phillips, Baer, Beer, Atala, and Rasmussen (1998).

Especially relevant to the present chapter are other, non delusional beliefs that are less dramatic, less widely known, but which are also commonly held by schizophrenics. These beliefs have been uncovered in studies of the phenomenology of schizophrenia, i.e., from first-person accounts from schizophrenics (Davidson, Stayner, and Haglund, 1998). One common finding is that even schizophrenics who are withdrawn and apparently indifferent to others describe themselves as being intensely lonely and haunted by the beliefs that they are "nothing" and that they will never have positive, affectionate relationships with others. They commonly believe that will not be accepted or welcomed into regular social groups and communities and that they are not full, regular members of the human race. They often believe that others see them as mental patients and not as full human beings with unique personalities and sensitivities. They believe that they will not be taken seriously when they talk. They believe that they are excessively sensitive to stimulation and cues during interpersonal encounters and live in fear that they will be overwhelmed and unable to respond. They believe that they will be unable to cope with even minor signs of negative affect or disapproval from others, and that they are unable to manage the demands and complexities of relationships. They often believe they have lost their identities, and that they will never be able to properly control their thoughts and behaviors.

**Personality Disorders**
Personality is typically defined as stable patterns of thoughts, feelings and behaviors that persist across time and situations. To a large degree, our personalities are defined by our stable beliefs, assumptions, and automatic thoughts about ourselves and others. In the DSM-IV-TR, personality disorders (PDs) are described as rigid, extreme, and maladaptive constellations of otherwise normal personality traits. It therefore directly follows that there will be prototypical sets of beliefs associated with each of the ten official PDs (Davidson, 2008). The most thorough elucidation of the beliefs associated with particular PDs was provided by Beck, Freeman, and Associates (1990) and by Beck, Freeman, Davis, and Associates (2004). The lists of beliefs for each PD are rather long and so the beliefs associated with just two PDs will be provided here. Common beliefs in the narcissistic PD include: "I am a very special person;" "Since I am so superior, I am entitled to special treatment and privileges;" "I don't have to be bound by the rules that apply to other people;" "It is very important to get recognition, praise, and admiration;" "If others don't respect my status, they should be punished;" "Other people should satisfy my needs;" "Other people should recognize how special I am;" "It is intolerable if I am not treated with respect or do not get what I am entitled to;" "Other people do not deserve the admiration or riches that they get;" "People have no right to criticize me;" "No one's needs should interfere with my own;" "Since I am so talented, people should go out of their way to promote my career;" "Only people as brilliant as I am understand me," and "I have every reason to expect grand things" (Beck et al., 1990, pp. 361-362).

Common beliefs in the antisocial PD include: "I have to look out for myself;" "Force or cunning is the best way to get things done;" "We live in a jungle and the strong person is the one who survives;" "People will get at me if I do not get them first;" "It is not important to keep promises or honor debts;" "Lying and cheating are OK as long as you don't get caught;" "I have been unfairly treated and am entitled to get my fair share by whatever means I can;" "Other people are weak and deserve to be taken;" "If I don't push other people, I will get pushed around;" "I should do whatever I can get away with;" "What others think of me doesn't really matter;" "If I want something, I should do whatever is necessary to get it;" "I can get away with things so I don't need to worry about bad consequences;" "If people can't take care of themselves, that's their problem" (Beck et al., 1990, p. 361). Lists of common beliefs involved in the paranoid, schizoid, schizotypal, avoidant, dependent, histrionic, borderline, and obsessive-compulsive PDs were also provided by Beck et al. (1990, 2004).

## DYNAMICS OF NORMAL AND PATHOLOGICAL BELIEF SYSTEMS

We believe that a dynamical, complex systems approach can provide a general model that encompasses the transition from a normal set of beliefs to worldviews that typify various pathologies. There has been much previous interest in dynamical models of mental states, including those associated with psychopathology (Mandell and Selz, 1995; Tschacher and Kupper, 2007). Examples include work on depression (Heiby et al., 2003), bipolar disorder (Gottschalk et al., 1995), post-traumatic stress (Horowitz, 1986), alcohol abuse (Hufford et al., 2003), schizophrenia (Paulus et al., 1996), and antisocial behavior in children and adolescents (Granic and Patterson, 2006). This previous research has focused on nonlinear, dynamical trends in diverse constellations of symptoms with particular focus on variations in the severity of *symptoms* across time. This work is certainly valuable, but we believe it is also important to focus on the development of maladaptive belief systems. A proper understanding of the dynamical patterns that underlie the cognitive domain of human functioning may prove useful in understanding the dynamical patterns that also exist in the symptoms that have been the focus of previous work.

### The Forging of a Worldview

We use the term 'worldview' to refer to one's internal model of the world. We want to clarify that by this term we mean much more than a collection of isolated memories, beliefs, attitudes, bits of knowledge, and so forth. The concept of a worldview encompasses the manner in which one navigates memories and bits of knowledge, weaving narratives with them, and thereby understanding and interacting with the world. We noted that a worldview is self-regenerating, or autopoietic, meaning that through the interactions amongst its parts it forges its structure as a whole. We also noted that it has been proposed that a worldview constitutes the fundamental structural unit of cultural evolution akin to the organism in biological evolution (Gabora, 1998, 2004, 2008). It changes through self-triggered thought as well as individual and social learning, and cultural evolution occurs when the structure of one worldview crystallizes, in whole or in part, in the mind of another individual.

The worldview of a normal adult is integrated as evidenced by our ability to prioritize activities of different sorts, combine information from different domains (as in analogical reasoning), adapt views and actions to new circumstances, frame new experiences in terms of previous ones, solve problems using peripheral cues, and formulate plans of action that reflect the specifics of a situation. Some of the contents of a worldview are acquired through social learning, e.g. imitation, social facilitation, and so forth. Most other worldview contents are clumped together as the product of individual learning. This includes items obtained through perceptual processes as well as those obtained through contemplation or learning processes that are not socially mediated. We can also distinguish between elements of the worldview that are assimilated more or less 'as is', and elements that undergo extensive modification in an attempt to frame it in one's own terms, adapt it to the unique structure of one's beliefs, or 'put one's own spin on it' (Gabora, under revision).

**The Worldview as Self-organizing and Self-mending**
A worldview is able to obtain and maintain this integrated structure because it exhibits a natural tendency for self-organization. It finds structure spontaneously through interactions amongst its parts. Narrative structures such as scripts and stories emerging from sequences of experiences, and attitudes emerging from specific beliefs, and these structures being in dynamic flux over time. Previous states influence present states, which influence future states. Internal feedback mechanisms, in combination with external influences, encourage dynamics within beliefs systems.

A worldview is self-mending in the sense that one is inclined to resolve threats to its congruity or integrity (Gabora, 1999; Heider, 1958; Greenwald et al., 2002; Osgood and Tannenbaum, 1955; Piaget, 1970). Much as a wounded organism may spontaneously heal, if we encounter something that challenges our knowledge and beliefs (e.g. a script plays out in an unexpected way, someone does something out of character, or an experimental result is different from what was predicted by theory), we experience cognitive dissonance and spontaneously attempt to weave a story, fortify our knowledge, or revise our beliefs to accommodate the challenge (Festinger, 1957).

A worldview is additionally self-regenerating in the sense that an adult shares concepts, ideas, attitudes, stories, and experiences with children (and other adults), influencing little by little the formation of other worldviews. Each worldview takes shape through the influence of many others, though some, such as those of parents and teachers, will predominate. The children expose fragments of what was originally the adult's worldview to different experiences, different bodily constraints, and thus sculpt unique internal models of the relation of self to world. One can also say that a worldview is autopoietic because the whole is reconstituted through the interactions of the parts (Maturana and Varela, 1980). One memory can evoke another, which evokes another, and so forth, and each time a memory is evoked it can be expressed, and thereby influence other worldviews.

**Cognitive Flexibility and the Structure of Associative Memory**
In order to see how the proclivity for self-triggered thought emerges, we look briefly to the structure of associative memory. We take as a starting point some fairly well established characteristics of memory. Human memories are encoded in neurons that are sensitive to ranges (or values) of subsymbolic microfeatures (Churchland and Sejnowski, 1992; Smolensky, 1988). For example, one might respond to a particular shade of red, or the quality of being shrewd, or quite likely, something that does not exactly match an established term (Mikkulainen, 1997). Although each neuron responds maximally to a particular microfeature, it responds to a lesser extent to related microfeatures, an organization referred to as coarse coding. Not only does a given neuron participate in the encoding of many memories, but each memory is encoded in many neurons. For example, neuron A may respond preferentially to lines of a certain angle (say 90 degrees), while its neighbor B responds preferentially to lines of a slightly different angle (say 91 degrees), and so forth. However, although A responds maximally to lines of 90 degrees, it responds somewhat to lines of 91 degrees. The upshot is that storage of an item is distributed across a cell assembly that contains many neurons, and likewise, each neuron participates in the storage of many items (Hinton, McClelland, and Rumelhart, 1986). Thus, the same neurons get used and re-used in different capacities, a phenomenon referred to as neural re-entrance (Edelman, 1987). Items stored in overlapping regions are correlated, or share features. Memory is said to be content addressable; there is a systematic relationship between the state of an input and the place it gets encoded. As a result, episodes stored in memory can thereafter be evoked by stimuli that are similar or 'resonant' in some (perhaps context-specific) way (Hebb, 1949; Marr, 1969).

The fact that memory is distributed and content-addressable is critically important to cognitive flexibility and the ability to escape rigid patterns. If it were not distributed, there would be no overlap between items that share sub-symbolic microfeatures, and thus no means of forging an association between them. If it were not content-addressable, associations would not be meaningful. Content addressability is why the entire memory does not have to be searched or randomly sampled; it ensures that one naturally retrieves items that are relevant. Content addressability also facilitates the activation of one item by another that is related to it in a rarely noticed but useful or appealing way. Recall that if the regions in memory where two distributed representations are encoded overlap, then they share one or more microfeatures. They may have been encoded at different times, under different circumstances, and the correlation between them never explicitly noticed. But the fact that their distributions overlap means that some context could come along for which this overlap would be relevant or useful, and cause one to evoke the other. Content addressability also means that there are as many routes to a reminding event as there are microfeatures by which they overlap; i.e. there is plenty of room for typical as well as atypical associations. It is because the region of activated memory locations falls midway between the two extremes—not distributed and fully distributed—that one can generate a stream of coherent yet potentially creative thought (Gabora, 2002). The more detail with which items have been encoded in memory, the greater their potential overlap with other items, and the more retrieval routes for creatively forging relationships between what is currently experienced and what has been experienced in the past, and thereby 'breaking out of a rut'.

**Navigating the 'Realm of Potentiality'**
We get ideas, make decisions, and plan courses of action, keeping an eye to what we expect and want from the future. Surrounding all present situations and events there can be said to be a 'halo' of possible future events, possible ways in which the present will unfold into the future. This can be referred to as the 'realm of potentiality' (Gabora and Aerts, 2005, 2007) or 'adjacent possible' (Kauffman, 2008). When one's conception of what is possible in the short term and long term is broad, and one is able to envision a multitude of ways in which a situation could unfold, we say that the situation is in a *potentiality state* with respect to the state of the worldview. When one's conception of the adjacent possible 'shrivels' and one's worldview no longer provides new ways of envisioning how a situation could unfold, we say that the situation is in an *eigenstate* with respect to the state of the worldview.

A narrowing over time of one's conception of what is immediately possible is natural and perhaps in some respects adaptive; there is no need to prepare oneself for the vast number of ways a situation could unfold if it almost always unfolds in a certain predictable way. However, a narrowing in one's conception of the adjacent possible surrounding the event can be associated with a belief structure that causes one to act in ways that reinforce particular trajectories and contribute to cognitive and behavioral rigidity.

**The Crystalization of Distorted Belief Structures**
The solidifying of a structure of beliefs is in many respects useful, for it automates and thus decreases effort needed for habitual acts. The solidification process makes use of the distributed, content-addressable, coarse-coded nature of memory, and the capacity for associative recall. Daily situations evoke responses, including all one's current thoughts and feelings about a situation, and particular regions of memory are activated. Items previously encoded to neurons activated by a situation provide 'ingredients' for the next thought. This next thought is slightly different from the one that preceded it, so it activates and retrieves from a slightly different region, and so forth. The interaction between the developing belief structure and the worldview in turn alters the worldview, actualizes some aspect of it that was before merely potential, and changes what is subsequently potential for it, i.e. how it will proceed *from there* to generate contexts. In this way, the belief structure is reiteratively reinterpreted or *redescribed* (Karmiloff-Smith, 1992) from various real or imagined perspectives or *internally-generated contexts*.

By internally generating contexts for the belief structure, elements external to the current subject of thought leak into one's conception of it. Eventually, a belief structure can be said to be in an *eigenstate* with respect to context provided by the worldview, i.e. no longer subject to change. The worldview does not continue to provide new contexts because the problem and worldview are no longer dissonant. The belief reached a state where it is consistent with the worldview, and the worldview has reached a state where it can incorporate the belief. Of course, if the worldview that guides this assimilation process is biased, the mental operations that transform the belief will distort it in such a manner as to strengthen and perpetuate the misguided worldview.

What causes impressions to organize themselves into a mental model of reality that is distorted? Belief systems, perhaps especially those associated with psychopathology, likely have two kinds of external input: (a) from the social and physical environments, and (b) from affective and biological factors within the person that are nevertheless external to the belief systems themselves. Specific beliefs are likely not continuously present in consciousness, but there are probably reliable patterns of temporal variation in maladaptive beliefs within each person. Furthermore, once an individual embarks on a train of thought, the direction and conclusions of the trains may become rigidly consistent and rapid over time. Attractors cause trains of thought with initially varying starting points to result in the same, repeating end states. It is also assumed that slight changes in internal or external control factors may cause dramatic changes in system dynamics and behavior.

**Interpersonal Processes in the Solidification of Belief Structures**
Cognitive and interpersonal processes are presumably closely intertwined as belief systems develop. Emerging beliefs about oneself and the world are initially experienced as tentative working hypotheses that require further investigation and confirmation. Yet the very methods that individuals naturally use to test their suspicions are known to be biased in ways that create and produce confirming evidence (Andrews, 1989; Carson, 1982; Snyder, 1984; Swann, Rentfrow, and Guinn, 2003). Beliefs have a tendency to constrict the range of possible behaviors in a situation. The behaviors that are emitted on the basis of particular beliefs then, in turn, naturally constrain or restrict the range of possible responses from others. The responses provided by others then tend to confirm the original beliefs that generated and skewed the whole process. If I think that I am a likeable person and that you are friendly, then I am likely to display friendly behaviors towards you that will prompt friendly responses from you. Your friendly responses will then serve to confirm my original hunches about your friendliness. Developing beliefs thus rapidly become solidified via the self-fulfilling prophecies that are involved in how beliefs are tested. The course of events is taken as evidence for the correctness of the original beliefs, which typically occurs without the realization of how the course of events was skewed by the original beliefs.

There is evidence that individuals engage in such self-fulfilling prophecies with imagined others (Andrews, 1989), perhaps especially once beliefs begin to solidify. As imagined or anticipated interpersonal scenarios become scripted, expectations about interactions become stronger, which further constricts how interactions unfold. A variety of well-established cognitive self-fulfilling processes can occur either on their own, or simultaneously with interpersonal processes, that further solidify the developing belief system. These include selective attention, selective interpretation, and selective recall (Mathews and Macleod, 2005; Swann et al., 2002).

Self-concepts and worldviews eventually solidify as a result of these cognitive and interpersonal processes. Indeed, one's personality is largely defined by one's recurring patterns of beliefs and by the behaviors and affective reactions that are generated by one's beliefs. Personality becomes difficult to change because belief systems and worldviews are naturally self-confirming. The flexibility and wide range of 'adjacent possibles' that existed before beliefs systems developed are gradually reduced. Individuals who are well-adjusted tend to have beliefs and behaviors that are more flexible than those of individuals who are less well-adjusted. But presumably anyone's range of thoughts and behaviors may be narrower and more maladaptive than they realize. Individuals are more likely to be happy and well-adjusted if they retain some flexibility and if their particular patterns of rigidity in thought and behavior do not produce aversive outcomes for themselves and others. Unfortunately, this is not always the case.

## IMPLICATIONS OF OUR MODEL AND ANSWERS TO PERENNIAL QUESTIONS

The above descriptions of the nature of beliefs systems and worldviews, of the nature of memory, of how one idea leads to another, and of how belief systems solidify, are all useful in explaining the basic phenomenon that is the focus of this chapter: How once-normal individuals develop deep, maladaptive convictions that are seemingly impervious to refuting evidence, and that are closely intertwined with psychiatric symptoms. The maladaptive beliefs that are associated with disorders are not present at birth. They are, instead, discovered and developed through basic, normal, universal, cognitive processes. We have brains that naturally weave and organize coherent beliefs systems from the wide array of everyday information inputs. Our brains are naturally drawn to inconsistencies in our belief systems and we attempt to resolve them. When our memories are searched during these self-mending efforts, our brains provide us with familiar bits of information that, at least initially, we can weave and piece together in novel, creative (for us) ways.

As our various beliefs become more idiosyncratically integrated and coherent, they are increasingly experienced as correct and real. Our very efforts to test the veracity of our beliefs tend to provide further confirming information. The range of alternative beliefs about ourselves and others shrivels. We slowly seal ourselves into self-made corners that seem absolutely real and true. Other people may seem foolish for not sharing our views. The entire trajectories in the development of belief systems are dynamic and complex. The primary differences between those with and without psychiatric problems in this view, are in the contents and rigidity of the belief systems, and not the basic processes.

**Maladaptive Creativity**

It could be said that the sets of maladaptive beliefs in the disorders described above are the products of creative minds. Creativity is required to fashion the coherent, integrated networks of beliefs from initially disconnected elements and experiences. Once these networks are in place, they become so automatically and routinely activated that the range of adjacent possibles is severely limited. Thought patterns become more clearly rigid. At this point, our natural, creative, self-mending and integrative tendencies still exist, but they may begin to work against us. Our brains now creatively twist what should be dissonant information into confirming information. "I clearly see that I really am fat, even though these people have hospitalized me for being too thin."

**The Pervasiveness of Disorders**

Our natural capacities for developing mental models of reality and for problem-solving leave us with the potential to develop maladaptive beliefs systems. This provides a partial answer to a vexing, perennial problem: Why are psychological disorders so pervasive and persistent? It is difficult to argue that so many forms of maladaptation were specifically selected by evolutionary forces. Neurological defects should not be common and should not persist through generations. (The psychopathic exploitation of others may be an exception.) Instead, what is likely inherited are not defects, but the basic architecture for problem-solving and for fashioning coherent mental models of the world.

The end results of our integrating, self-mending thoughts may sometimes be distinctly maladaptive, but the ability to process a truly wide range of possible inputs and to generate creative integrations and solutions was likely a central reason why the capacity for self-triggered thought, and thereby an integrated worldview, evolved. The same brain capacities that lead us to create novel technologies and artwork can also lead us seriously astray. Normality and abnormality can thus be seen as belonging in the same cognitive framework.

**Sources of Belief Contents**
If the primary differences between individuals with and without psychiatric problems are in the contents of their beliefs systems, and not the basic processes that generated the beliefs, then it is important to consider where the peculiar contents come from. We have described how maladaptive beliefs can be fashioned from diverse inputs, and there must certainly be differences between normal and abnormal people with regards to the nature of the inputs. There appear to be two broad, interrelated categories. First, the family, peer, and community experiences are known to be quite different for those with and without problems. Real-world risk factors result in distinct inputs and basic beliefs about oneself and others.

A second, related important source of input is one's own affective and physiological states. Energy levels, brain chemistry, and how one's body feels undoubtedly influence the thoughts one has about oneself and others. When the input from internal states is consistent over time, our brains must weave this input into a network with other inputs.

On subsequent occasions, when physiological states may be more positive than is usually the case, individuals may be led us back to chronic darker views because of the mountains of evidence and the network of automatic assumptions that have become so available and entrenched.

**Self-Defeating and Persistent Beliefs**
We can also speculate as to why beliefs systems sometimes become abnormally rigid and resistant to change, even when they are self-defeating and continually produce negative outcomes. Our above descriptions of how and why worldviews tend to solidify at least partially explains the common, normal levels of rigidity in belief systems. We suspect that excessive rigidity develops from the cognitive processes that are used to test the veracity of beliefs, especially when individuals are threatened or stressed. When beliefs are questioned and doubts begin to arise, perhaps as a result of disconfirming feedback from others, individuals search the current situation and their memories in an effort to determine what is real and true. But when it is the experience of negative affect that initiates this process, then individuals may be particularly likely fall back and activate their existing belief networks. When threatened, it is easier and more comforting to find evidence for what one already knows and believes than to seriously question existing assumptions, and entertain the possibility that reality is dramatically different from what has long been assumed to be the case. Strong pulls may be exerted on others to provide reactions and feedback that confirms one's threatened but preferred views (Andrews, 1989; Kiesler, 1996). As individuals repeatedly pursue these methods of mending inconsistencies in their belief networks, confirming what they already know may become their sole source of comfort and reassurance. What people conclude each time their beliefs networks are activated and then confirmed may be grim and ultimately self-defeating, but it is also very familiar, real, and reassuring (Andrews, 1989; Swann et al., 2003). The 'adjacent possibles' for more adaptive beliefs are increasingly limited and the change that would be required is too great and discomforting.

**Inconsistent Beliefs**

Gaps and inconsistencies in beliefs systems generate negative affect and self-mending cognitive activity. Dissonance reduction efforts are often but not always successful. Un-resolvable conflicts between beliefs may result in aversive symptoms, such as those in post-traumatic stress disorder (PTSD).

In other words, while anxiety and negative affect may be generated by beliefs about specific threats, as in phobias, they can also be generated by conflicts between networks of beliefs that are frequently activated but that result in dramatically different and incompatible conclusions about the nature of reality.It is also important to note that some disorders, such as bipolar disorder and the borderline personality disorder, are characterized by alternation between dramatically different psychological states. Those with such disorders presumably have rich, frequently activated networks of beliefs for each of the states that they are prone to.

Borderline patients are sometimes deeply loving, accepting, and appreciative of close others. But other times their distrust, anger, and rage at the same close others can be powerful and highly destructive. It is as if they have two strongly contrasting sets of internal working models of themselves and others, and both sets are experienced as absolutely real. In these cases, both the actual states and the shifts between states are problematic for themselves and others.

**Diagnostic Comorbidity and Disorder Prototypes**

A final feature of our model is that it provides a partial explanation for the often high levels of diagnostic comorbidity and for the many patients who seem to have blends of disorders and yet who are not highly prototypical of any single disorder. It seems only natural that some of the same maladaptive beliefs will be evident in different disorders. The beliefs associated with schizophrenia likely encourage the development of beliefs that are also seen in depression. The networks of beliefs in schizophrenia and depression are adjacent possibles. The result is individuals who may meet the criteria for both disorders. The same applies to other forms of comorbidity. Also relevant are patients who are clearly disordered but who do not meet the criteria for specific disorders. An example is that fact that "Personality Disorder Not Otherwise Specified" is the mostly commonly diagnosed PD in clinical settings. This likely occurs because belief systems are not required to develop in conformity with particular templates. Instead, creative, organizing, and integrative cognition is applied to varying inputs and experiences, which can result in highly idiosyncratic belief systems, some of which are maladaptively rigid and extreme.

**Causal Influences and Possibilities for Change**

In this chapter we have described the dynamics of beliefs processes that we suspect occur in a variety of disorders. Our model is descriptive. Causal associations between beliefs and disorders presumably sometimes exist and operate in both directions. The associations may take at least three different forms (Klein, Wonderlich, and Shea, 1993). First, rigid belief systems may operate as a predisposing or vulnerability factor for psychopathology, and psychopathology may operate as a predisposing or vulnerability factor for rigid beliefs. In the latter case, abnormal affective and physiological states generate cognitive inputs that becomes integrated with other input into coherent belief systems. Second, even when beliefs do not play a predisposing or causal role in disorders, they may affect the course or expression of disorders (the "pathoplasty" model). Thoughts about oneself and others can result in affective states, behaviors, and interpersonal experiences that can modify the course of a disorder. Similarly, the existence of a disorder may alter the development of belief systems. Individuals discover new information about themselves as a result of having a disorder. A third possibility is the complication or scar hypothesis. In this case, belief systems may be fundamentally altered in enduring ways by the experience of a disorder and the effects may persist once the disorder itself disappears. In summary, beliefs may sometimes play a causal role in disorders, and they are also influenced by disorders. The directions of influence probably vary across both individuals and disorders.

Regardless of the origins of rigid beliefs systems, once they exist, they are likely problematic for the individual and for those with whom the individual interacts. The rich, readily activated networks of available thoughts can make the individual very certain, stubborn, and exasperating to be with, particularly when their beliefs are ultimately self-defeating. A belief network that has become highly woven, integrated, and repeatedly activated does not suddenly disappear, even when other symptoms begin to fade. Rigid maladaptive belief systems likely fade only slowly, and only when the individual fashions alternative belief networks that are frequently activated and that seem more convincingly real than the pre-existing beliefs. Relapses may be common as individuals tentatively shift their mental models of reality.

## SUMMARY AND CONCLUSIONS

In this chapter we presented a general dynamical process model of belief systems in psychiatric disorders. It is proposed that for a broad range of different disorders the basic underlying cognitive processes involved in the transition from beliefs are normal and adaptive to beliefs that are rigid, extreme, and maladaptive are of a similar nature. We further propose that they are all side-effects of the fact that the web of understandings a human mind weaves about the world is not just complex, dynamic, self-organizing, but also self-regenerating, self-perpetuating, and thereby alive and evolving not just at the biological organismic level but at a second, cultural level. The ability to weave an internal model of the world, or worldview, stems from the specifically human capacity for self-triggered stream of thought, in which one thought triggers another, which triggers another, and so forth. Self-triggered thought enables our original or raw impressions of the world to be assimilated into a growing, ever-changing mental model of reality, or worldview. In so doing we weave a more-or-less coherent web of understanding, which makes the human mind particularly adept at planning and prioritizing, drawing analogies, adapting behavior to the specifics of situations, and so forth. However, this transformation process can also cause impressions to become increasingly distorted. The more distorted the developing worldview, the greater the extent to which newly forming impressions and beliefs become distorted as they are made sense of by this worldview. The relevant thought trajectories are self-confirming, and are considered to underlie the corresponding trajectories in symptoms. In contrast with previous work, the model proposed here focuses on underlying mechanisms and it provides an evolutionary basis for the widespread susceptibility to psychiatric symptoms and disorders without the problematic claim that such disorders were selected by evolutionary forces. The model thereby incorporates both normality and abnormality in the same framework.